\documentclass[aps, pra, amsmath,amsfonts,reprint,superscriptaddress,longbibliography]{revtex4-1}
\usepackage{amsmath,amssymb,amsfonts}
\usepackage{color}
\usepackage{natbib}
\usepackage{graphicx}
\usepackage{dcolumn}
\usepackage{bm}
\usepackage{color,soul}
\usepackage{lipsum}

\usepackage[usenames]{xcolor}

\begin{document}

\title{Time-varying metasurfaces for broadband spectral camouflage}

\author{Mingkai Liu}
\affiliation {Nonlinear Physics Centre, Research School of Physics, Australian National University, Canberra ACT 2601, Australia}
\author{Alexander B. Kozyrev}
\affiliation {Advanced Technology, Mission Systems, 
Collins Aerospace, 400 Collins Road, Cedar Rapids, IA 52498 USA}
\author{Ilya V. Shadrivov}
\email{ilya.shadrivov@anu.edu.au}
\affiliation {Nonlinear Physics Centre, Research School of Physics, Australian National University, Canberra ACT 2601, Australia}

\begin{abstract}
The possibility of making an object invisible for detectors has become a topic of considerable interest over the past decades. Most of the studies so far focused on reducing the visibility by reshaping the electromagnetic scattering in the spatial domain. In fact, by manipulating the electromagnetic scattering in the time domain, the visibility of an object can also be reduced. Importantly, unlike previous studies on phase-switched screens and time-varying metasurfaces, where the effect is narrow band due to the dispersive resonance, for microwave frequency range, we introduce a broadband switchable metasurface integrated with PIN diodes. The reflection phase of the metasurface can be changed by approximately $\pi$ over a fractional bandwidth of 76$\%$. By modulating the metasurface quasi-randomly in the time-domain, the incident narrow-band signal is spread into a white noise-like spectrum upon reflection, creating a spectral camouflage. The broadband feature of the proposed time-varying metasurface can provide practical insight for various applications including radar stealth and ultrawide-band wireless communication.

\end{abstract}


\maketitle

\section{Introduction}

We see an object when the electromagnetic signal, either emitted or scattered, is detected and distinguished from the background noise. To reduce the visibility of an object, we need to minimize the wave energy that propagates in the directions of the detectors. Traditionally, this has been realized by absorbing the incoming electromagnetic wave using absorbing coating materials, or redirecting the scattered wave via a deliberate design of the object's geometry or the morphology of the coating material~\cite{rao2002integrated,lynch2004introduction}. 

Recent developments in metamaterials and metasurfaces have shown their tremendous capabilities in manipulating the propagation and scattering of electromagnetic waves~\cite{engheta2006metamaterials,glybovski2016metasurfaces,genevet2017recent,ding2017gradient}, extending previous studies on frequency-selective surfaces~\cite{mittra1988techniques,munk2005frequency}, blazed gratings~\cite{aoyagi1976high,feng2007polarization} and phased-array antennas~\cite{berry1963reflectarray,pozar1993analysis}. These developments also offer alternative concepts and platforms for reducing the visibility of objects. The most intriguing approach is to utilize the concept of transformation optics to construct a metamaterial cloak that can bend the incoming electromagnetic wave around the object such that the scattered waveform seems unperturbed~\cite{pendry2006controlling,schurig2006metamaterial,li2008hiding}. Meanwhile, more practical approaches based on metasurfaces have been developed, including ultra-thin absorbers ~\cite{landy2008perfect,ding2012ultra,sun2011extremely,sun2012broadband} and diffusers~\cite{gao2015broadband,chen2017coding,modi2017novel} that can reduce the radar cross section (RCS) of the object. The device performance can be further tuned or switched using {\em tunable metasurfaces}~\cite{fan2015dynamic,liu2017ultrathin,hajian2019active}. Being able to fabricate {\em flexible metasurfaces} further extends the scope for potential applications~\cite{iwaszczuk2012flexible,chen2017coding}.

 \begin{figure}[t!]
\includegraphics[width=1.0\columnwidth]{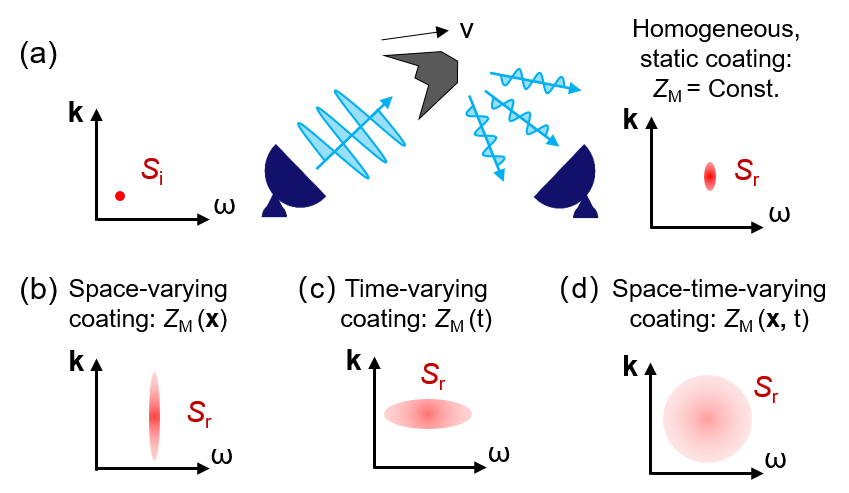}
\caption{ (a) Schematic of radar detection of a moving object. The coordinate of $(\bold{k},\omega)$ represents the power density distribution of the signal $S$ in the momentum domain $\bold{k}$ and the spectral domain $\omega$. $S_i$ and $S_r$ are the incident and the back-scattered signal, respectively. For a moving object with a homogeneous and static coating layer, represented by the metasurface impedance $Z_M$, the power density of the back-scattered wave experiences a Doppler shift. (b) For an object coated with randomly distributed space-varying coating $Z_M(\bold{x})$, the back-scattered power density is spread in the momentum domain. (c) For an object coated with time-varying coating $Z_M(t)$, the scattered power density is spread in the spectral domain as sidebands. (d) The power density at sideband frequencies can be further suppressed using space-time-varying coating $Z_M(\bold{x},t)$.  \label{fig:concept}}
\end{figure}

In a typical radar detection process, the object's  scattered signals are analyzed in the frequency domain, and its distance and motion information can be retrieved from the Doppler spectral features~(Fig.~\ref{fig:concept}a)~\cite{chen2006micro}. Existing methods for reducing objects' visibility are mainly based on manipulating the wave in the spatial domain. For example, the momentum of the propagating incident wave is largely converted into fast-decaying evanescent waves after the wave interacts with metasurface absorbers. While a metasurface diffuser can spread the momentum of the incoming wave into a broad spectrum~(Fig.~\ref{fig:concept}b), thus reducing the power density of the reflected signal.  

From the space-time analogy for electromagnetic waves, it is natural to ask if an object's visibility can also be reduced by manipulating the scattering signal in the time domain. Just like a space-varying metasurface diffuser, a time-varying metasurface can, in principle, achieve the same goal by spreading the power of scattered waves over a broad spectral domain~(Fig.~\ref{fig:concept}c). More importantly, for radar detection, if the scattered signal is severely modulated before it is detected, it becomes highly challenging to retrieve the object's information reliably without the prior knowledge of the modulation. For example, Doppler illusion (signature distortion) can be realized by shifting the scattered frequency to one or more new frequencies via a periodic modulation of the surface impedance~\cite{liu2018huygens,zhao2018programmable,zhang2018space}.

Indeed, this concept is closely related to spread-spectrum communication~\cite{peterson1995introduction}, and its potential for stealth technology has been explored in the context of phase-switched screens~\cite{tennant1997reflection,tennant1998experimental,chambers2002general,chambers2004phase,xu2016improved,wang2018synthetic,wang2019spread}. The concept can be realized by modulating the reflection phase of a Salisbury screen between $\pi$ and $0$ to spread the power efficiently from the carrier frequency to the sidebands~\cite{tennant1997reflection,chambers2004phase}. Recent development in time-varying metasurfaces and antenna arrays further enriches the physics and applications of such time-varying systems~\cite{liu2018huygens,salary2018time,caloz2019spacetime}, including efficient frequency conversion and dynamic beam shaping~\cite{liu2018huygens,zhao2018programmable,zhang2018space,lee2018linear}, nonreciprocal wave controls~\cite{shaltout2015time,hadad2015space,taravati2016mixer,zang2019nonreciprocal}, and new architectures of wireless communications~\cite{zhao2018programmable,tang2019wireless1,han2019large,tang2019wireless2}.

 However, existing design of phase-switched screens and tunable metasurfaces can only offer a $\pi$ phase shift within a narrow frequency band due to the dispersive nature of the resonance~\cite{tennant1998experimental,zhao2018programmable,wang2019spread}. From the practical point of view, it is highly desirable that the surface can offer a $\pi$-phase switch over a broad frequency band. In this work, we show a design of switchable metasurface with integrated PIN diodes, which can achieve a reflection phase tuning of around $\pi$ over a wide frequency band. By modulating this broadband switchable metasurface in a quasi-random fashion, the incident narrow band signal can be transformed into a broadband noise-like spectrum upon reflection, creating a spectral camouflage. Our designed metasurfaces offer a wider bandwidth and larger angular tolerance compared to previous works on phase-switched screens and time-varying metasurfaces. The reflected power level at the carrier frequency can be suppressed to below $-10$ dB over a fractional bandwidth of 76\%, and the performance can be maintained over a wide range of incident angles. In addition, the presented mechanism allows for latency-free dynamic RCS manipulation~\cite{song2019broadband}, which can enable target features transformation in arbitrary fashion and make identification of the target by conventional radar systems significantly harder.

\section{Results} 

 \begin{figure}[t!]
\includegraphics[width=1.0\columnwidth]{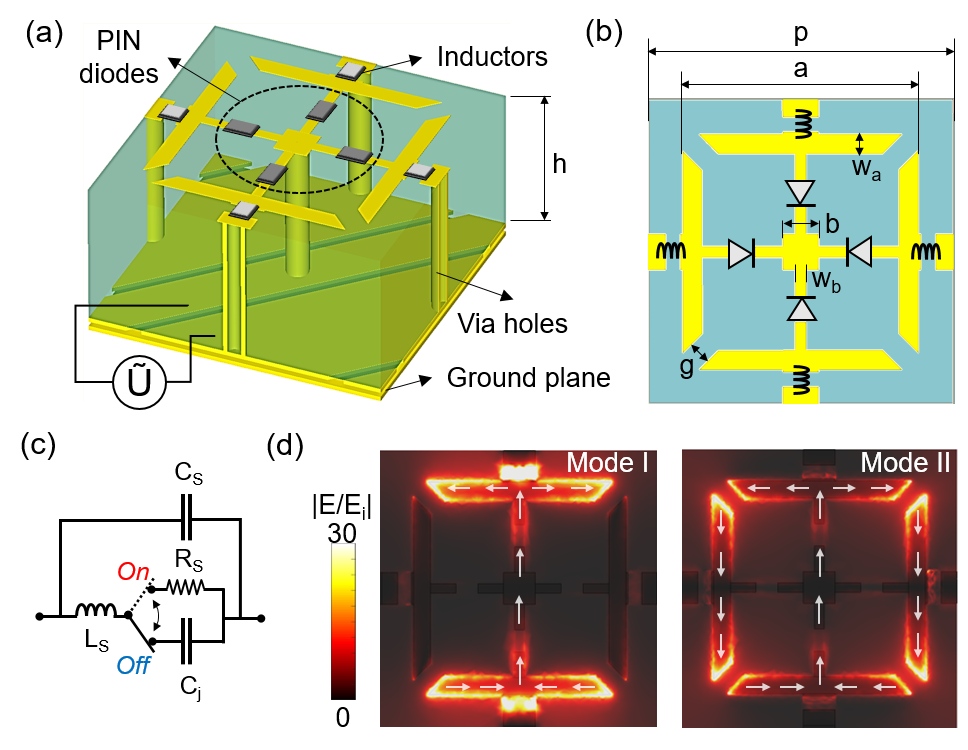}
\caption{ (a) Schematic of the broadband switchable metasurface. The top metasurface contains PIN-didoes, which can be switched by the bias signal $\tilde{U}$. The additional metal plate below the bias electrodes serves as the ground plane. The substrate material is Rogers RO3003, with a thickness $h=6.35 $mm. The inductors serve as low-pass filters for the modulation signal and have an inductance of 100 nH. The diameter of the via-holes is 1 mm. (b) The geometries of the top metasurface (units in mm): $p=13$, $a=10$, $b=1.5$, $w_a=0.8$, $w_b=0.45$, $g=1$. (c) The equivalent circuit of the PIN-diode (SMP1345-079LF, Skyworks Solutions Inc): $L_s=0.7$~nH, $R_s=2~\Omega$, $C_j=0.15$~pF, $C_S=0.18$~pF. (d) The electric field distributions of the two modes supported by the meta-unit. The field amplitude has been normalized to the incident field $E_i$. The white arrows indicate the flow of the surface current.    \label{fig:design}}
\end{figure}

The metasurface unit (dubbed meta-unit below) design is based on a metallic resonator printed on a Rogers RO3003 substrate, backed by a ground plane, as shown in Fig.~\ref{fig:design}a. We combine full-wave and circuit simulations to design the meta-unit using CST Microwave Studio, with periodic boundary conditions in the lateral directions and a plane-wave excitation in the normal direction. The four-fold rotational symmetry of the top layer pattern is chosen to achieve polarization-independent response, and their detailed geometries are given in the caption to Fig.~\ref{fig:design}b. PIN-diodes (SMP1345-079LF, Skyworks Solutions Inc)~\cite{xu2016dynamical} are employed as the tunable elements, with their equivalent circuit model shown in Fig.~\ref{fig:design}c. The top metallic patterns are connected to the bottom electrodes through the via holes such that the diodes can be modulated with a voltage signal $\tilde{U}$. The flat metal ground plate is attached to the bottom electrodes, with an insulation layer between them (Fig.~\ref{fig:design}a). The meta-units can support \emph{two resonant modes} (see Fig.~\ref{fig:design}d): one has a dipole-like current distribution (mode I), and the other has a toroidal like current distribution (mode II). 

\begin{figure}[t!]
\includegraphics[width=1.0\columnwidth]{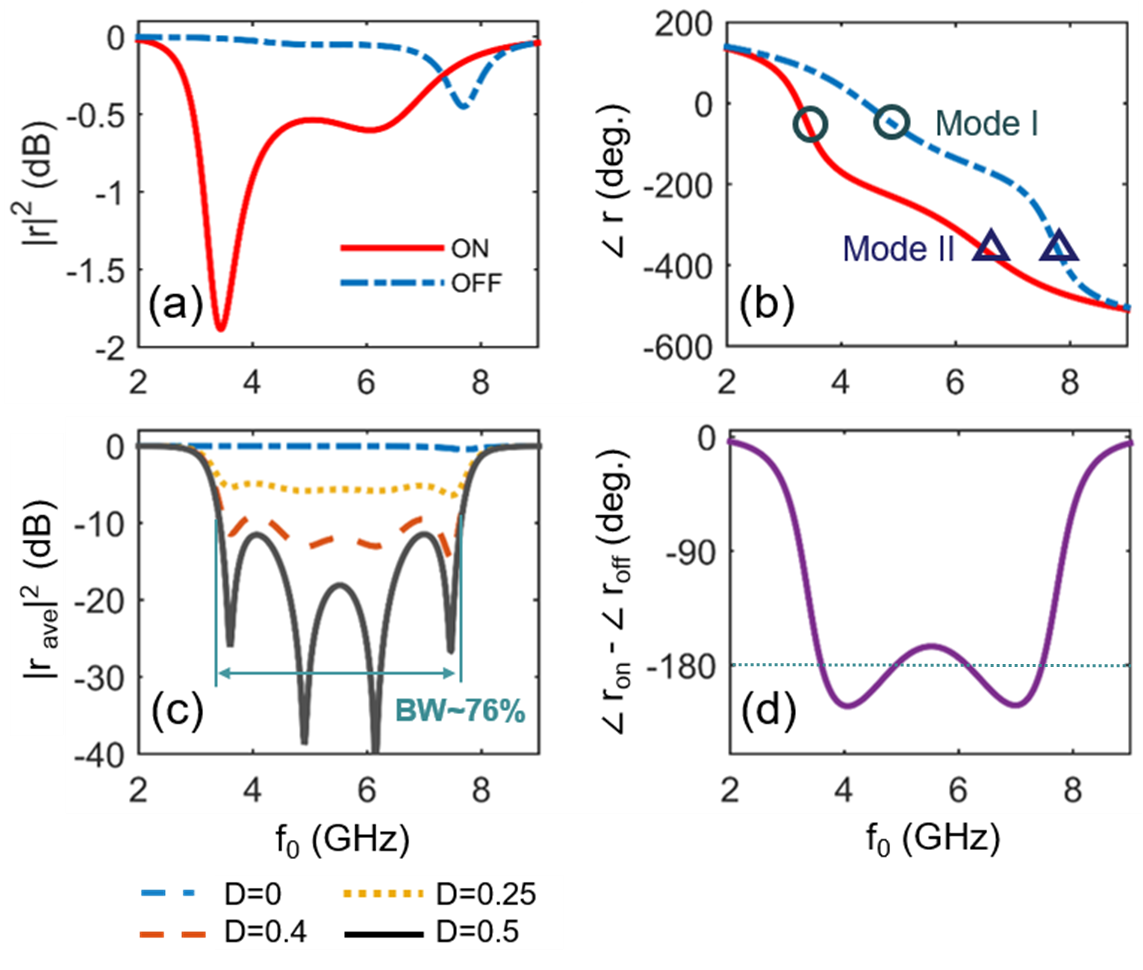}
\caption{ (a) and (b) Linear reflection spectra of the metasurface in ``on" and ``off" states under a plane-wave excitation at normal incidence. The circles and triangles indicate the frequencies of the two resonant modes shown in Fig.~\ref{fig:design}d. The corresponding reflection phase difference between states ``on" and ``off" is shown in (d). (c) The time-averaged reflectance at the \emph{carrier frequency} $f_0=\omega_0/(2\pi)$ for different modulation duty cycle $D$. A $10$ dB change in the reflection power can be achieved over a fractional bandwidth of 76\% by varying the duty cycle $D$. \label{fig:effect}}
\end{figure}

The resonant frequencies of the two modes shift when the PIN diodes are switched from state ``On" to state ``Off", as shown by the linear reflection spectra in Fig.~\ref{fig:effect}a and~\ref{fig:effect}b. The reflection amplitude is not unit due to the resonant absorption in the metasurface. Assuming that the metasurface is switched between the ``On" and ``Off" states in a time-varying fashion, the spectra of the incident signal $s_i(t)$ and the modulated reflected signal $s_r(t)$ can be obtained via a Fourier transformation
\begin{eqnarray}
s_{i}(\omega)&=&\mathcal{F}[s_i(t)],\\
s_{r}(\omega)&=&\mathcal{F}[r(t)s_i(t)].   
\end{eqnarray}
In our study, the bandwidth of modulation is much smaller than the resonant bandwidth of the metasurface and the incident carrier frequency $\omega_0$. Thus, the time-varying reflection coefficient can be approximated with $r(t)=r_{\rm on}(\omega_0)$ when the voltage signal $\tilde{U}(t)=1$, and $r(t)=r_{\rm off}(\omega_0)$ when $\tilde{U}(t)=-1$. The time-averaged reflection coefficient at the carrier frequency $\omega_0$ is then given by
\begin{equation}\label{eq:r_ave}
 r_{\rm ave}(\omega_0)=\frac{s_{r}(\omega_0)}{s_{i}(\omega_0)}=Dr_{\rm on}(\omega_0)+(1-D)r_{\rm off}(\omega_0),
\end{equation}
where $D$ is the percentage of time (duty cycle) when the metasurface is in the state ``On". When $D\approx 0.5$, a $\pi$ phase difference between $r_{\rm on}$ and $r_{\rm off}$ allows them to interfere destructively in time such that the time-averaged amplitude $|r_{\rm ave}(\omega_0)|\rightarrow 0$. 

To maximize the bandwidth of operation and the tuning range of reflection, the geometries of the meta-units are judiciously optimized such that the phase difference between $r_{\rm on}$ and $r_{\rm off}$ can be maintained around $\pi$ over a broad frequency range (Fig.~\ref{fig:effect}d). For a duty cycle $D=0.5$, in order to suppress the reflection at the carrier frequency to below $-10$~dB, i.e. $|r_{\rm ave}(\omega_0)|^2<0.1$, we found that the relative phase change should satisfy $143^{\circ}<|\angle r_{\rm on}- \angle r_{\rm off}|< 217^{\circ}$. Figure~\ref{fig:effect}c shows the time-averaged reflection spectrum of $r_{\rm ave}(\omega_0)$ for different values of duty cycle $D$. We clearly see that the time-averaged reflectivity can be tuned substantially by changing the duty cycle $D$. When $D$ increases from 0 to 0.5, a $10$~dB change of the reflection can be achieved over a 76\% fractional bandwidth, from around 3.4 GHz to 7.6 GHz (Fig.~\ref{fig:effect}c), which is significantly larger than previous designs based on a single resonance~\cite{tennant1998experimental,zhao2018programmable}. The positions of the four dips shown in the curve of $D = 0.5$ directly correspond to the frequency points where a $\pi$ phase difference occurs, as shown in Fig.~\ref{fig:effect}d. This broadband feature can be maintained even under a large angle of incidence with an improved design of the bias electrodes. For more detailed discussion on the performance under oblique incidence, see Appendix~\ref{sec:appendixA}.

 The broadband feature of our metasurface allows the object to be hidden more effectively from a wide range of detectors including frequency-agile radars. Ultimately, an object becomes untraceable if the Doppler spectral features are submerged in the background noise. To achieve this goal, the metasurface needs to be modulated with a quasi-random signal, such that the energy of the back-scattered wave can be redistributed evenly across a wide spectral range to create a white-noise like sideband spectrum within the filter bandwidth of the radar system. When most of the energy from the carrier frequency is converted to the sidebands, from energy conservation, one can estimate that the average power level of the generated sideband noise should satisfy the following approximate relation:
 \begin{equation}\label{eq:sidebandpower}
 |s_r(\omega\neq\omega_0)|^2\approx {\Gamma_i}/{\Gamma_m}|s_i(\omega_0)|^2,
 \end{equation}
 where $\Gamma_i$ and $\Gamma_m$ are the bandwidths of the incident and modulated reflected signals.

\begin{figure}[t!]
\includegraphics[width=1\columnwidth]{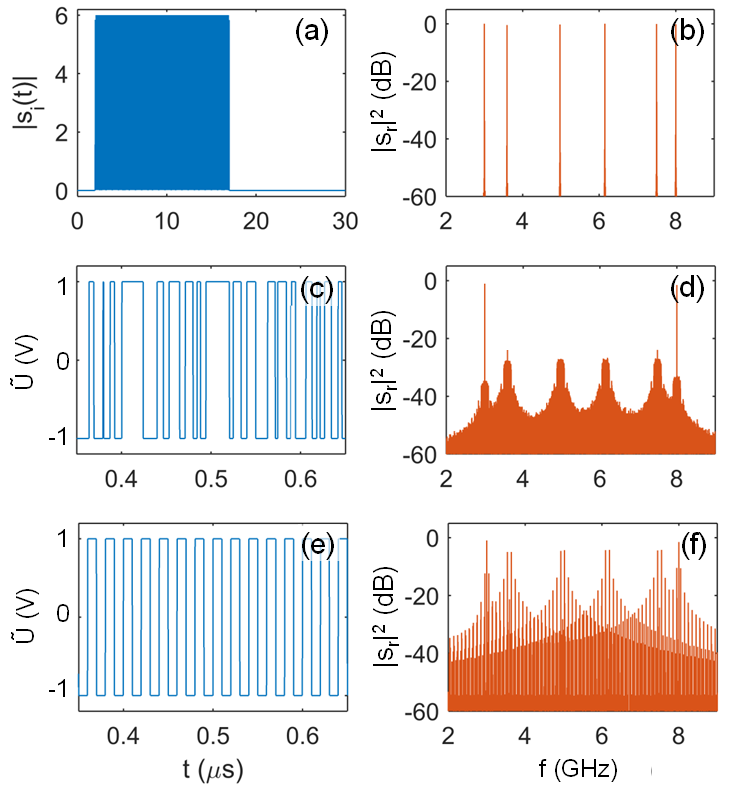}
\caption{ (a) The incident signal of a radar pulse train that has a period of pulse repetition of 30~$\mu$s and a pulse duration of 15~$\mu$s. To show the broadband effect, six carrier frequencies (3.0 GHz, 3.59 GHz, 5.00 GHz, 6.15 GHz, 7.5 GHz, 8.0 GHz) are employed simultaneously. (b) The spectrum of the reflected signal from a static metasurface. (c) A section of the bandwidth-limited ($<100$ MHz) random square-wave signal used to modulate the metasurface. (d) The spectrum of the reflected signal from the randomly modulated metasurface. (c) A section of the periodic (50 MHz) square-wave modulation signal and (d) the spectrum of the reflected signal from a periodically modulated  metasurface. \label{fig:testing}}
\end{figure}

As an example, we employ a bandwidth-limited ($<$ 100 MHz) quasi-random binary signal to modulate an infinitely large homogeneous metasurface.  The incident signal $s_i(t)$ is a rectangular pulse train:
\begin{eqnarray}
s_i(t)=\sin(\omega_0t)\mathbf{rect}(t/T_p),
\end{eqnarray}
which has a pulse duration of $T_p=15~\mu$s (Fig.~\ref{fig:testing}a). 
To show the broadband features of the metasurface, six carrier frequencies [$\omega_0/(2\pi)=$ 3.0 GHz, 3.59 GHz, 5.00 GHz, 6.15 GHz, 7.5 GHz, 8.0 GHz] are employed simultaneously within the 15~$\mu$s-long pulse. The lowest and highest frequencies are located outside the -10 dB operational frequency band shown in Fig.~\ref{fig:effect}c, while the rest four are all within the operational band. When there is no modulation applied, the intensity of the reflected signal from the metasurface is high, and only the carrier frequencies can be observed (Fig.~\ref{fig:testing}b). Once a quasi-random modulation is applied~(Fig.~\ref{fig:testing}c), the energy of the original narrow band signal is spread into a noise-like spectrum centered around the carrier frequencies (Fig.~\ref{fig:testing}d). For the four frequencies within the operational band, the reflected power levels at the carrier frequencies are suppressed to below $-24$ dB, and the average level of the sideband noise is around $-30$ dB. In contrast, the reflection power levels of the two frequencies outside the operational band remain very high due to the low energy conversion efficiency from the carrier waves to the sidebands. As a comparison, when the metasurface is modulated periodically (Fig.~\ref{fig:testing}e), the reflected spectrum shows distinct sideband features at discrete frequencies centered around the carrier frequencies (Fig.~\ref{fig:testing}f). Within the -10 dB operational band, the peak power level of the sidebands is around 20 dB larger than the one obtained using quasi-random modulation. 

Note that for the \emph{carrier frequencies}, the achievable minimal reflection is limited by the non-ideal $\pi$ phase difference between $r_{\rm on}$ and $r_{\rm off}$, as indicated by Eq.~(\ref{eq:r_ave}). While for the \emph{sideband frequencies}, a further reduction of the power level can be realized if the bandwidth of the random modulation is increased (i.e. to increase $\Gamma_m$), or a longer pulse is employed (i.e. to decrease $\Gamma_i$), as can be inferred from Eq.~(\ref{eq:sidebandpower}). In order to achieve a high-performance spectral camouflage, i.e. the reflection power level is suppressed significantly (ideally below the background noise level) at both the carrier and sideband frequencies, it requires that the phase switch is close to $\pi$ and that the modulation bandwidth is much larger than the incident one. 

However, in practical applications, there could be situations where the modulation frequency is comparable to or even smaller than the incident signal bandwidth. In these scenarios, although the reflection power level can not be reduced significantly, the reflected Doppler signal is still severely distorted in both amplitude and phase, mimicking a white noise spectrum. Such a distortion can impose a significant challenge for radar signal processing if the detection side does not possess the prior knowledge of the modulation sequence. For more detailed studies on the stealth application of phase-switched screens, see Refs.~\cite{xu2016improved,wang2018synthetic}.

\begin{figure}[t!]
\includegraphics[width=1.0\columnwidth]{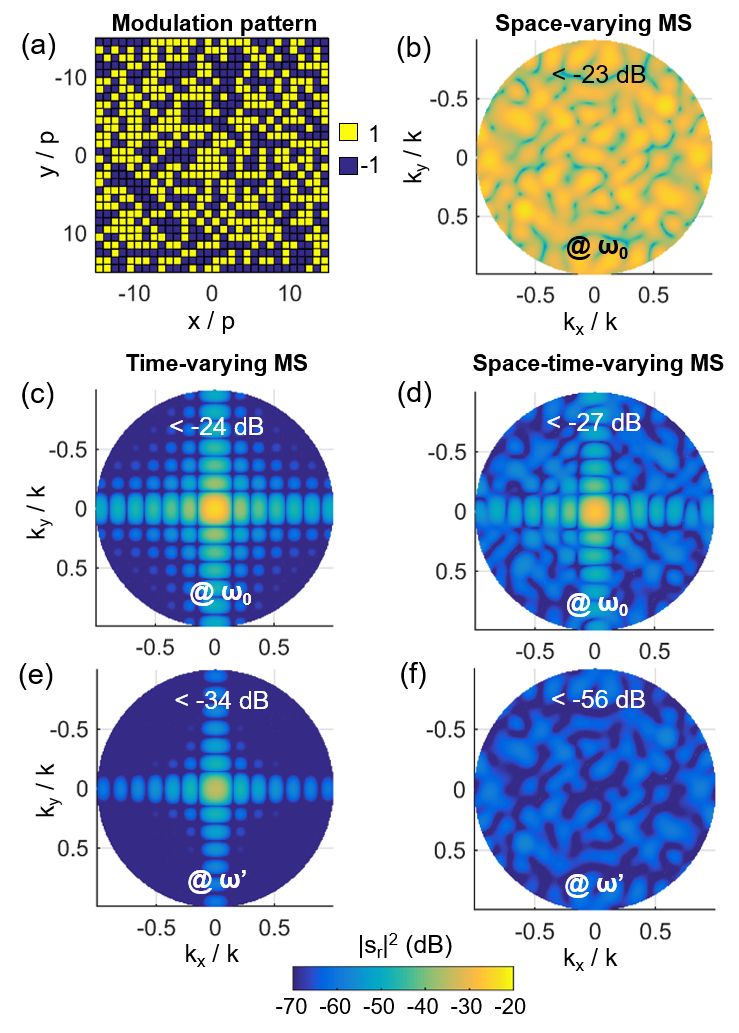}
\caption{ (a) Space-varying pattern of a finite-size (30 by 30 units) metasurface. (b) The scattering power distribution in the momentum space from a static space-varying metasurface (MS) at the carrier frequency $\omega_0/(2\pi) = 5$~GHz. The scattering power distribution at (c) the carrier frequency $\omega_0/(2\pi) = 5$~GHz and (e) one of the sidebands $\omega'/(2\pi) = 5.08$~GHz, when only time-varying modulation is applied to a homogeneous array. The corresponding responses under space-time-varying modulation are shown in (d) and (f). \label{fig:momentumspect}}
\end{figure}

Finally, we extend the analysis to finite arrays and analysis their performance for RCS control. In Fig.~\ref{fig:momentumspect}, we examine the performance of scattering suppression in a space-varying static metasurface (Fig.~\ref{fig:momentumspect}b), a homogeneous time-varying  metasurface (Fig.~\ref{fig:momentumspect}c and e), and a  space-time-varying metasurface (Fig.~\ref{fig:momentumspect}d and f). The size of the array is 30 by 30 units, and they are excited evenly with a normally incident plane wave. The momentum distribution of the back-scattered wave for the finite array can be approximated by
\begin{equation}
s_{r}(k_{x},k_{y},\omega)=\mathcal{E}(k_{x},k_{y})\times\sum_{mn}s_{r,mn}(\omega)e^{i(k_{x}x_{mn}+k_{y}y_{mn})} ,
\end{equation}
where $\mathcal{E}(k_x, k_y)$ is the radiation pattern of a single meta-unit (element factor), and $s_{r,mn}(\omega)$ denotes the local contribution of the back-scattered signal from meta-unit $(m,n)$. Here, we assume that the change of mutual interaction in the finite and inhomogeneous array is small, and that the local response of the meta-unit can still be approximated by the response from the periodic array.    

 The random distribution of sites ``1" and ``-1" shown in Fig.~\ref{fig:momentumspect}a indicates that the meta-units at these two sites are oppositely biased. This random pattern is employed for space-varying and space-time-varying modulation. To achieve this pattern, one can simply reverse the direction of the PIN-diodes at sites ``1" with respect to sites ``-1", such that the reflected signals from the two sites are modulated in an anti-phase fashion.  

The back-scattered power density distributions $|s_r|^2$ shown from Fig.~\ref{fig:momentumspect}b to f have been normalized to the peak power density from a homogeneous static metasurface, which is highly reflective. 
For a space-varying static metasurface, the level of back-scattered power density at the carrier frequency $\omega_0/(2\pi)=5$~GHz is suppressed to below $-23$ dB by redistributing the scattered power evenly in the momentum domain (Fig.~\ref{fig:momentumspect}b). For a homogeneous time-varying metasurface, only specular reflection is expected, and the momentum distribution of the back-scattered signal are similar at the carrier frequency (Fig.~\ref{fig:momentumspect}c) and the sidebands (Fig.~\ref{fig:momentumspect}e). The small side lobes are due to the finite size of the metasurface. Nevertheless, a $-24$ dB reduction at the carrier frequency is obtained, consistent with the result of an infinitely large time-varying array shown in Fig.~\ref{fig:testing}d. 

Interestingly, when space and time modulations are applied simultaneously, the peak reflected power at the carrier frequency only reduces slightly to $-27$ dB (Fig.~\ref{fig:momentumspect}d). This is because the time-varying modulation signal employed in Fig.~\ref{fig:momentumspect}c and d is already nearly optimal, even though the modulations at sites ``1" and ``-1" are out of phase, their time-averaged contributions at the carrier frequency are largely the same, with minor phase difference. As a result, the space-varying pattern does not have a large impact on the carrier frequency. In contrast, the phases at sideband frequencies have a gauge freedom controlled by the modulation~\cite{liu2018huygens}. Therefore, the sidebands experience additional phase difference at sites ``1" and ``-1" due to the anti-phase modulation. As shown in Fig.~\ref{fig:momentumspect}f, when space-varying modulation is introduced, the scattered power at sidebands is further spread in the momentum domain and the peak power density is reduced to below  $-56$ dB.     

\section{Discussion and Conclusion}
The double resonance of the designed meta-units provide the physical foundation for the broadband operation. The reflection suppression at the carrier frequency is limited by the non-ideal $\pi$ phase change, as shown in Fig.~\ref{fig:effect}d. We expect that this problem can be minimized with more sophisticated designs, for example, by interleaving meta-units that have different resonant frequencies to compensate the dispersion. We also note that the current operational bandwidth is still smaller than the maximal bandwidth allowed for a static absorber that has the same thickness~\cite{rozanov2000ultimate}, and we believe a further expansion of the operation frequency band is possible when additional resonances are introduced.

In order to achieve the best performance, it's also important to minimize unnecessary losses. As shown in Fig.~\ref{fig:effect}c, since dynamic controls of radar cross section is an important feature of the time-varying approach, excessive loss will affect the dynamic range of the modulation. In addition, since loss can affect the amplitude of reflection around the resonance, when excessive loss exists, the reflection amplitudes in ``on" and ``off" states become highly imbalanced, and the broadband feature of the reflection suppression effect can also be affected.   

To conclude, we proposed a novel design of switchable metasurface at the microwave frequencies, which can provide a phase change around $\pi$ over a broad frequency band. We demonstrated numerically that by modulating the metasurface in a quasi-random fashion, the incident narrow band signal can be spread into a broadband white-noise like spectrum. We further examined the behavior in a finite array, showing that the back-scattered power at the sidebands can be further suppressed by applying a space-varying modulation. Our study can provide a practical insight for the development of broadband phase-shift screen, benefiting applications ranging from radar stealth to ultra-wideband wireless communication. By enabling multiple resonances, we show the pathway to achieve the maximum possible bandwidth in the time-varying metasurface system, which is also multi-functional as the same metasurface can be used to perform a variety of functions, such as frequency conversion, beam steering, filtering, etc., in addition to the exemplary spectral camouflage shown in this work. These functions can be switched on the fly, and such performance flexibility certainly can not be achieved in simple static absorbers. Such hardware sharing will ultimately lead to significant size and weight reduction in comparison to the systems where dedicated hardware is used for each function. The proposed concepts can be readily applied to other frequency ranges once the fast tunable metasurfaces are available in those frequency ranges. 

\section*{Author Contribution}
M.~Liu conceived the idea, performed the theoretical and numerical studies; A.~B. Kozyrev provided insights on potential applications and desired functionality of metasurface devices; I.~V.~Shadrivov supervised the project. M. Liu wrote the manuscript, with input from all coauthors.

\section*{Acknowledgements}
This work was supported by the Australian Research Council. The authors thank Prof. Christophe Caloz for useful discussions.

\appendix
\section{Design and performance under oblique incidence}\label{sec:appendixA}

\begin{figure}[t!]
\includegraphics[width=1.0\columnwidth]{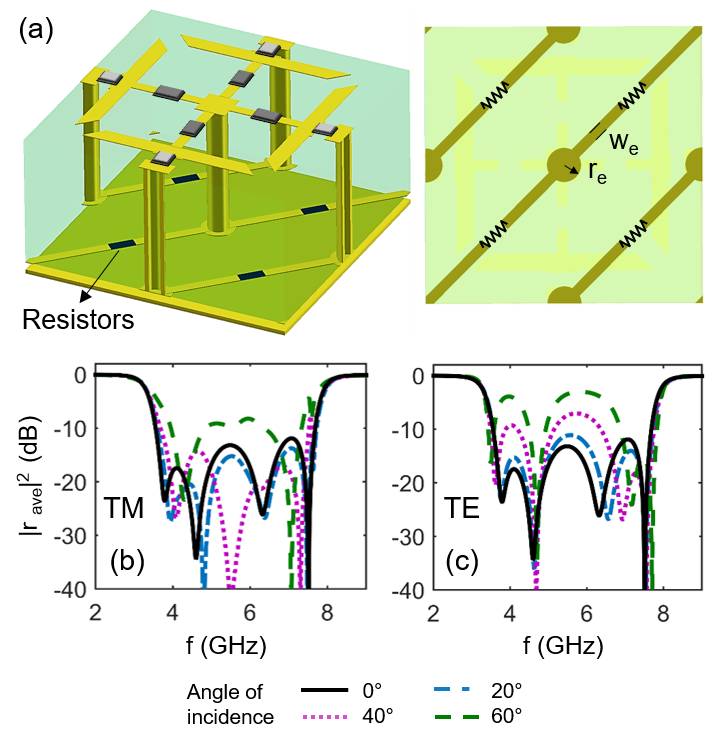}
\caption{ (a) Design of meta-unit with modified bias electrodes and additional resistors. $w_e=0.5$ mm, $r_e=0.8$ mm. (b) and (c) The time-averaged reflectance at the carrier frequency under different incident angles and polarizations. The modulation duty cycle $D=0.5$.  \label{fig:obliqe}}
\end{figure}

The design shown in Fig.~\ref{fig:design}a is optimized for the polarization-independent response under normal incidence. The performance under oblique incidence depends on the polarization and the angle of incidence. We found that the design shown in Fig.~\ref{fig:design}a is not the optimal for oblique incidence since additional sharp resonances exist when the metasurface is excited by the transverse magnetic (TM) polarization, damaging the broadband response. These resonances are surface modes supported by the network of vias and electrodes, which are dark modes at normal incidence. One way to suppress these undesirable sharp resonances is to introduce additional loss. 

As an example shown in Fig.~\ref{fig:obliqe}a, we integrate additional 15~$\Omega$ resistors to the bottom electrodes, while all other geometric parameters of the top layer remain the same as the ones shown in Fig.~\ref{fig:design}a. With the improved design of bias electrodes, we re-simulate the performance of the time-varying metasurface at oblique incidence for both transverse electric (TE) and transverse magnetic (TM) polarizations. For a modulation duty cycle $D=0.5$, the time-averaged reflectance spectra at the carrier frequency are shown in Fig.~\ref{fig:obliqe}b and c. Since the broadband response is mainly determined by the top-layer pattern, the resistors on the bottom electrodes have a minor impact and the spectral feature of broadband reflection suppression is largely the same as the one shown in Fig.~\ref{fig:effect}c. Remarkably, this broadband effect is quite robust and can be maintained even when the incident angle increases to around 40 degrees, thanks to the subwavelength size of the meta-unit ($\sim\lambda$/4, $\lambda$ is the central wavelength of the operation frequency band).

\bibliography{TMS_RCS}

\end{document}